\documentclass[reprint,superscriptaddress,amsmath,amssymb,aps,pre]{revtex4-1}

\usepackage{graphicx}
\usepackage{dcolumn}
\usepackage{bm}
\usepackage{color}
\usepackage{bbold}
\usepackage{soul}
\usepackage{epsfig,epic}
\newcommand{\newc}{\newcommand}
\newc{\beq}{\begin{equation}}
\newc{\eeq}{\end{equation}}
\newc{\kt}{\rangle}
\newc{\br}{\langle}
\newc{\beqa}{\begin{eqnarray}}
\newc{\eeqa}{\end{eqnarray}}
\newc{\longra}{\longrightarrow}

\providecommand*{\oneD}{\textsc{1d}}
\providecommand*{\twoD}{\textsc{2d}}
\providecommand*{\threeD}{\textsc{3d}}
\providecommand*{\fourD}{\textsc{4d}}

\providecommand*{\psslc}{\threeD{} phase-space slice}
\providecommand*{\psslcs}{\psslc s}

\makeatletter
\let\Hy@backout\@gobble
\makeatother

\begin{document}

\title{What is the mechanism of power-law distributed Poincar\'e
  recurrences \\ in higher-dimensional systems?}

\author{Steffen Lange}
\affiliation{Technische Universit\"at Dresden, Institut f\"ur Theoretische
             Physik and Center for Dynamics, 01062 Dresden, Germany}

\author{Arnd B\"acker}
\affiliation{Technische Universit\"at Dresden, Institut f\"ur Theoretische
             Physik and Center for Dynamics, 01062 Dresden, Germany}
\affiliation{Max-Planck-Institut f\"ur Physik komplexer Systeme, N\"othnitzer
Stra\ss{}e 38, 01187 Dresden, Germany}

\author{Roland Ketzmerick}
\affiliation{Technische Universit\"at Dresden, Institut f\"ur Theoretische
             Physik and Center for Dynamics, 01062 Dresden, Germany}
\affiliation{Max-Planck-Institut f\"ur Physik komplexer Systeme, N\"othnitzer
Stra\ss{}e 38, 01187 Dresden, Germany}

\date{\today}

\begin{abstract}
  The statistics of Poincar\'e recurrence times in Hamiltonian systems
  typically shows a power-law decay with chaotic trajectories sticking
  to some phase-space regions for long times. For higher-dimensional systems the mechanism of this
  power-law trapping is still unknown. We investigate trapped orbits of a generic \fourD{}
  symplectic map in phase space and frequency space and find that, in
  contrast to \twoD{} maps, the trapping is (i) not due to a hierarchy
  in phase space. Instead, it occurs at the surface of the regular
  region, (ii) outside of the Arnold web. The chaotic dynamics in this
  sticky region is (iii) dominated by resonance channels which reach
  far into the chaotic region: We observe (iii.a) clear signatures of
  some kind of partial transport barriers and
  conjecture (iii.b) a stochastic process with an effective drift
  along resonance channels. These two processes lay the basis for a future
  understanding of the mechanism of power-law trapping in
  higher-dimensional systems.
\end{abstract}

\pacs{PACS here}

\maketitle

Chaotic transport in Hamiltonian systems is of great importance in a
wide variety of applications, e.g., for predicting the stability of
celestial motion and satellites~\cite{MurHol2001,Cin2002,
  DaqRosAleDelValRos2016}, controlling the beams of particle
accelerators~\cite{DumLas1993,VraIslBou1997,RobSteLasNad2000,Pap2014},
and to describe the dynamics of atoms and
molecules~\cite{SchBuc2001,TodKomKonBerRic2005, GekMaiBarUze2006,
  PasChaUze2008, WaaSchWig2008, ManKes2014}. Generic Hamiltonian systems are not fully chaotic
but have a mixed phase space which also contains regular tori.
Close to these tori the chaotic transport is slowed down
considerably. This intermittent behavior is
characterized by the (cumulative) Poincar\'e recurrence statistics $P(t)$, the
probability that a chaotic orbit has not returned to an initial region
within time $t$. While for fully chaotic systems $P(t)$ usually
decays exponentially~\cite{BauBer1990, HirSauVai1999}, for mixed
systems the decay is much slower following a power law $P(t)\sim
t^{-\gamma}$~\cite{ChiShe1983, Kar1983}. This so-called power-law trapping or
stickiness entails dramatic consequences for the transport properties
in many systems, e.g., comets in the solar system~\cite{She2010b},
reactant lifetime in transition state theory~\cite{EzrWaaWig2009},
\textsc{DNA}~\cite{MazShe2015}, intramolecular energy redistribution~\cite{SetKes2012},
and microwave ionization of Rydberg
atoms~\cite{BucDelZakManArnWal1995, SchBuc2001,
  BenCasMasShe2000}.

Power-law trapping is well understood for Hamiltonian systems with two
degrees of freedom~\cite{KayMeiPer1984b, MeiOtt1985, RomWig1990, Zas2002, WeiHufKet2003, CriKet2008, AltTel2008,
  CedAga2013, AluFisMei2014, Mei2015}: In this
case the regular tori are barriers in phase space such that chaotic
orbits cannot cross them. In their vicinity so-called partial
transport barriers of the same dimension exist which can be crossed by
chaotic orbits, but with a small rate. A hierarchy of these partial
barriers governs the dynamics and causes the power-law trapping.

\begin{figure}[b]
\includegraphics{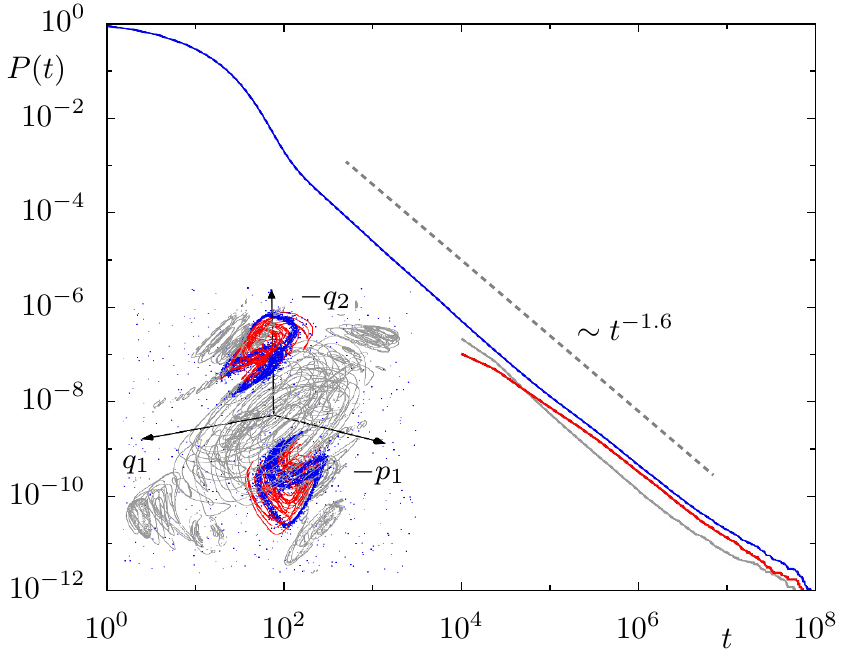}
\caption{\label{fig:power-law-trapping} Statistics of Poincar\'e
  recurrences $P(t)$ for the map~\eqref{eq:map} with initial region
  $q_1<0.1$ showing a power-law decay $P(t)\sim t^{-1.6}$. For
  $t>10^4$ the statistics (blue) is decomposed into orbits trapped at
  the dominant sticky region (red) and the remainder (gray). The inset
  shows the \psslc{} with $\vec{n}=(0,1,0,0)$, $D=0$ with a trapped
  orbit (blue points) and regular tori (red and gray rings).
  For a rotating view of the inset see the supplemental material \cite{supplementalvideos}.
}
\end{figure}

Power-law trapping is also observed in higher-dimensional
systems~\cite{DinBouOtt1990,AltKan2007,ShoLiKomTod2007b,She2010,SilManBeiAlt2015}, see
Fig.~\ref{fig:power-law-trapping} for an illustration. However, its
origin is unknown as the mechanism for two degrees of freedom cannot
be generalized: The distinctive feature of higher-dimensional systems
is that the regular tori have an insufficient dimension to be barriers
in phase space (i.e. the $f$-dimensional tori have at least two
dimensions less than the $(2f-1)$-dimensional energy surface for
$f\geq3$ degrees of freedom). Consequentially, any chaotic orbit can
get arbitrarily close to any regular torus by transport along
resonance channels of the so-called Arnold
web~\cite{Arn1964,Chi1979,Loc1999}. While the diffusion along a
channel is sometimes found~\cite{WooLicLie1990, Las1993,
  GuzLegFro2005} and often assumed~\cite{GuzLeg2013, EftHar2013} to be
normal, its detailed understanding is still
missing~\cite{Loc1999,EftHar2013,MesBazCinGio2014}. Generalizations of
partial barriers to higher-dimensional systems
exist~\cite{MarDavEzr1987,Wig1990,MacMei1992,ShoLiKomTod2007b} but
their relevance for the chaotic transport is unclear. The phenomenon
of power-law trapping can be explained in weakly coupled systems for
intermediate time-scales~\cite{AltKan2007, SilManBeiAlt2015}. Still, the generic
mechanism of power-law trapping in higher-dimensional systems remains
an open question.

In this paper we investigate the mechanism of power-law trapping in a
generic \fourD{} symplectic map, which corresponds to the lowest
dimensional Hamiltonian system for which regular tori are no barriers
in phase space. We find that the trapping is (i) not due to a
hierarchy in phase space, in contrast to \twoD{} maps. Moreover, it
occurs at the surface of the regular region, (ii) outside of the
Arnold web. We find that the chaotic dynamics in this sticky region is
(iii) dominated by resonance channels, which extend out of the Arnold
web far into the chaotic region: We demonstrate (iii.a) clear signatures
of some kind of partial transport barriers for the transport across
channels. We conjecture that (iii.b) a stochastic process with an
effective drift models the transport along a channel. Determining
which of these two processes is dominant should allow for unraveling
the mechanism of power-law trapping in higher-dimensional Hamiltonian
systems.

{\it System and Poincar\'e recurrence statistics.}--- To study the
power-law trapping for \fourD{} maps, we consider two coupled standard
maps~\cite{Fro1972}, $(p_1, p_2, q_1, q_2) \mapsto (p_1', p_2', q_1',
q_2')$,
\begin{equation}
\label{eq:map}
  q_i' = q_i + p_i \qquad p_i' = p_i - \frac{\partial
    V_i}{\partial q_i}(q_i') - \frac{\partial
    V_{12}}{\partial q_i}(q_1',q_2')
\end{equation}
with potentials $V_i = K_i/(4\pi^2) \cos(2\pi q_i)$ and coupling
$V_{12} = \xi_{12}/(4\pi^2) \cos(2\pi (q_1+q_2))$. Choosing the
nonlinearity parameters $K_1 = 2.25$, $K_2 = 3.0$ and a large coupling
parameter $\xi_{12}=1.0$, Eq.~\eqref{eq:map} represents a generic
\fourD{} symplectic map far from
integrability~\cite{RicLanBaeKet2014}. There is an elliptic--elliptic
fixed point at $(p_1, p_2, q_1, q_2)=(0,0,0.5,0.5)$ which
is surrounded by regular tori. For a \fourD{} map such regular tori
are two-dimensional and are organized around families of elliptic
\oneD{} tori which form a hierarchy~\cite{LanRicOnkBaeKet2014}. These
regular structures are embedded in a large chaotic region. We obtain the
statistics of Poincar\'e recurrences $P(t)$ for the map~\eqref{eq:map}
in a phase space $p_{1, 2} \in [-0.5, 0.5)$ and $q_{1, 2} \in [0, 1)$
with periodic boundaries and the initial region $q_1<0.1$, which
contains only chaotic dynamics. The resulting statistics $P(t)$ in
Fig.~\ref{fig:power-law-trapping} (solid blue line) exhibits a
power-law decay over several orders of magnitude. Chaotic orbits with
large recurrence times $t$ stick to the vicinity of different regions
of regular tori. As an illustration the \psslc{} (explained below) in
the inset in Fig.~\ref{fig:power-law-trapping} shows regular tori
(gray) and a chaotic orbit (blue) which sticks to a region of regular
tori marked in red. It turns out that this sticky region is dominantly
responsible for the power-law decay for $10^5<t<10^8$: The
decomposition of $P(t)$ for $t>10^4$ demonstrates that the fraction of
orbits sticking to this region (red line) is much bigger than the rest
(gray line). Thus, for the rest of the paper we focus on chaotic
orbits trapped in this dominant sticky region.

\begin{figure}
\includegraphics{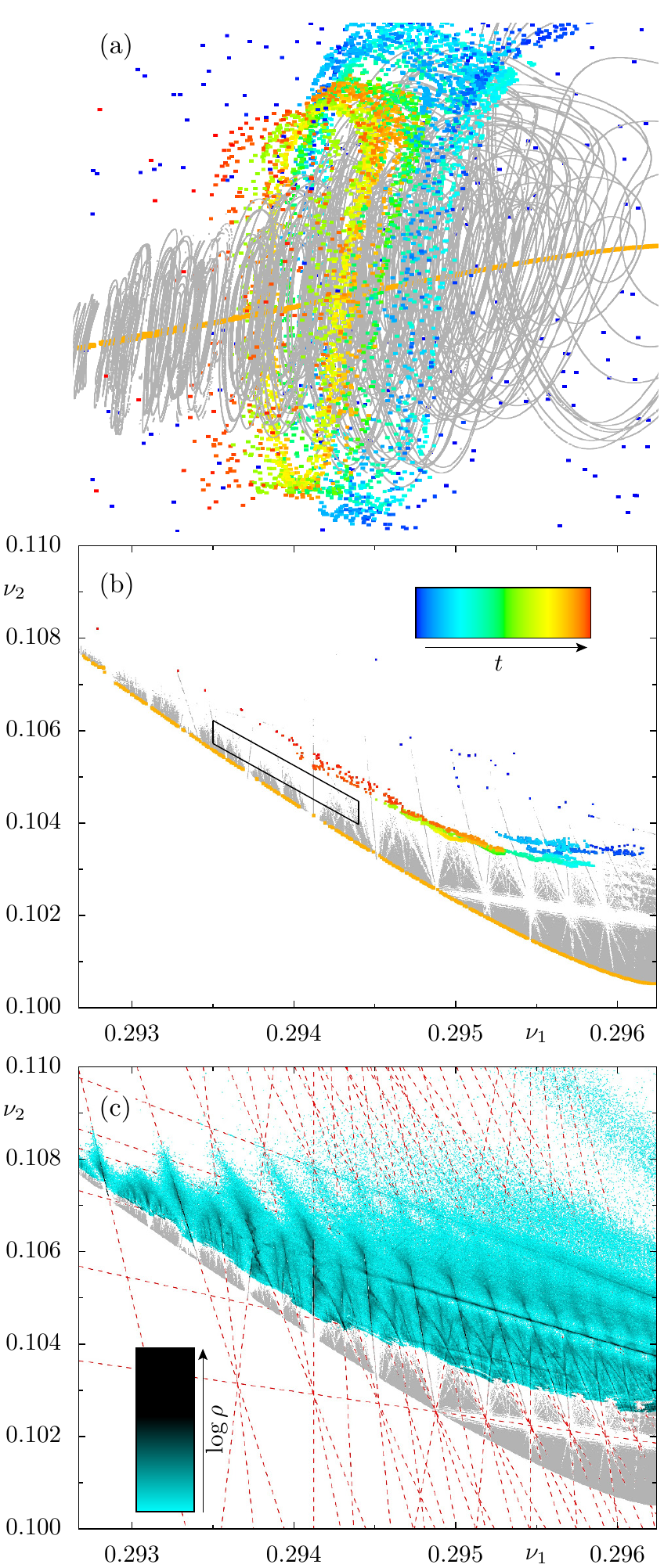}
\caption{\label{fig:freq-space-trapping} Dominant sticky region in (a)
  \psslc{} ($\vec{n}=(0.632,0.611,0.314,0.359)$, $D=0.322$) and (b),
  (c) frequency space with regular tori (gray (a) rings and (b), (c) points) and elliptic \oneD{}
  tori (orange). In (a), (b) a single trapped orbit is shown with the
  time $t \in [0, 1.7 \cdot 10^7)$ encoded in color (blue to
  red). In (c) the density $\rho(\nu_1,\nu_2)$ of frequencies, see (ii), is
  shown with resonance lines (red). White corresponds to $\rho=0$, turquois to black to $\rho \in [4\cdot 10^{-7}, 2\cdot 10^{-3}]$.
  For a rotating view of (a) see the supplemental material \cite{supplementalvideos}.
}
\end{figure}

{\it Visualization in phase space and frequency space.}--- In order to
analyze the mechanism of power-law trapping, we study the trapped
orbits in phase space, using \psslcs{}~\cite{RicLanBaeKet2014},
and frequency space. This is illustrated in
Figs.~\ref{fig:freq-space-trapping}(a) and (b), respectively, for a
typical chaotic orbit (same as in the inset of
Fig.~\ref{fig:power-law-trapping}) trapped in the dominant sticky region: For
a \psslc{} only points $\vec{u}=(p_1, p_2, q_1, q_2)$ within a thin
\threeD{} slice are displayed in the remaining three dimensions. The
slice is defined by $|\vec{u}\cdot\vec{n}-D|<10^{-4}$ with normal
vector $\vec{n}$ and distance $D$ to the origin given in the captions
of Fig.~\ref{fig:power-law-trapping} and
Fig.~\ref{fig:freq-space-trapping}. Phase-space objects typically
appear in such a slice with a dimension reduced by one. For instance
the regular \twoD{} tori appear as rings (gray) in
Fig.~\ref{fig:freq-space-trapping}(a) and the elliptic \oneD{} tori
appear as points (orange). Since elliptic \oneD{} tori occur in
one-parameter families around which the regular tori are
organized~\cite{LanRicOnkBaeKet2014}, the orange points form a line in
the center of the gray rings.

In order to relate trapped orbits to resonances, we complement the
\psslc{} with a frequency analysis~\cite{Las1990,
  BarBazGioScaTod1996,RicLanBaeKet2014}, see
Fig.~\ref{fig:freq-space-trapping}(b): For this, each \twoD{} torus is
associated with its two fundamental frequencies $(\nu_1, \nu_2) \in
[0, 1)^2$, i.e.\ a point in frequency space. The frequencies $(\nu_1,
\nu_2)$ describe the angular dynamics on the torus and are computed
from $\Delta t=4096$ iterations of an orbit on the torus. In the
frequency space in Fig.~\ref{fig:freq-space-trapping}(b) regular tori
(gray points) are organized above the family of elliptic \oneD{} tori
(orange points). The region of the regular tori is interrupted by
channels around resonance lines $m_1:m_2:n$, on which the frequencies
fulfill $m_1 \cdot \nu_1 + m_2 \cdot \nu_2 = n$ with integers $m_1,
m_2, n$. These so-called \emph{resonance channels} are accessible to
chaotic dynamics. Their network within the region of regular tori
is referred to as \emph{Arnold web}. A chaotic orbit can be displayed
in frequency space by decomposing the orbit into segments of length
$\Delta t$ and numerically assigning frequencies to each
segment~\cite{MarDavEzr1987,Las1993}.

{\it (i) Trapping not due to hierarchy.}--- A typical trapped orbit is shown in
Figs.~\ref{fig:freq-space-trapping}(a) and (b) with its points colored
according to iteration time: The orbit enters the sticky region
(blue), is trapped (bright blue to orange), and leaves the region
(red). The \psslc{} and the frequency space demonstrate that the
chaotic orbit is not trapped deep in the hierarchy: In
Fig.~\ref{fig:freq-space-trapping}(a) the trapped orbit is spread over
the surface of the regular structure without clustering on smaller
scales. In contrast, trapped orbits in \twoD{} maps spend more and
more time on finer and finer phase-space scales due to the hierarchy
of partial barriers. Furthermore, in
Fig.~\ref{fig:freq-space-trapping}(b) the orbit extends over several
resonances, whereas an orbit being trapped in the deeper levels of the
island-around-island hierarchy of a \fourD{} map would have
frequencies on a single resonance line or at a junction of resonances~\cite{LanRicOnkBaeKet2014}. We
estimate that the trapped orbits spend on average just $1 \%$ of their
time deeper in the hierarchy, e.g. near those regular tori which appear
on resonance lines in Fig.~\ref{fig:freq-space-trapping}(b).

{\it (ii) Trapping outside of Arnold web.}--- We analyze all trapped orbits
with $10^5<t<10^9$ using their density $\rho(\nu_1, \nu_2)$ in
frequency space, see Fig.~\ref{fig:freq-space-trapping}(c). The
density is computed from $61600$ trapped orbits on a grid with
resolution $\Delta \nu=5\cdot 10^{-6}$ and shown in logarithmic scale.
The density is zero in the resonance channels of the Arnold web. This
means that on the considered time scale the power-law trapping occurs
outside of the Arnold web.

{\it (iii) Relevance of resonance channels.}--- Remarkably, the
density in Fig.~\ref{fig:freq-space-trapping}(c) exhibits pronounced
peaks along resonance lines, some of which are indicated by red dashed
lines in the background. These resonance lines extend out of the
Arnold web into the chaotic region. For a chaotic orbit to exhibit
resonant frequencies it has to be confined around a resonance line at
least for the time interval $\Delta t$ of the frequency analysis. This
would be the case within resonance channels in the Arnold web, where they are
surrounded by regular tori. The high densities in
Fig.~\ref{fig:freq-space-trapping}(c) demonstrate that these resonance
channels still dominate transport outside the Arnold web, where they
are no longer surrounded by regular tori. Consequently, the chaotic
transport in the sticky region can locally be decomposed into
transport across and along resonance channels.

\begin{figure}
\includegraphics{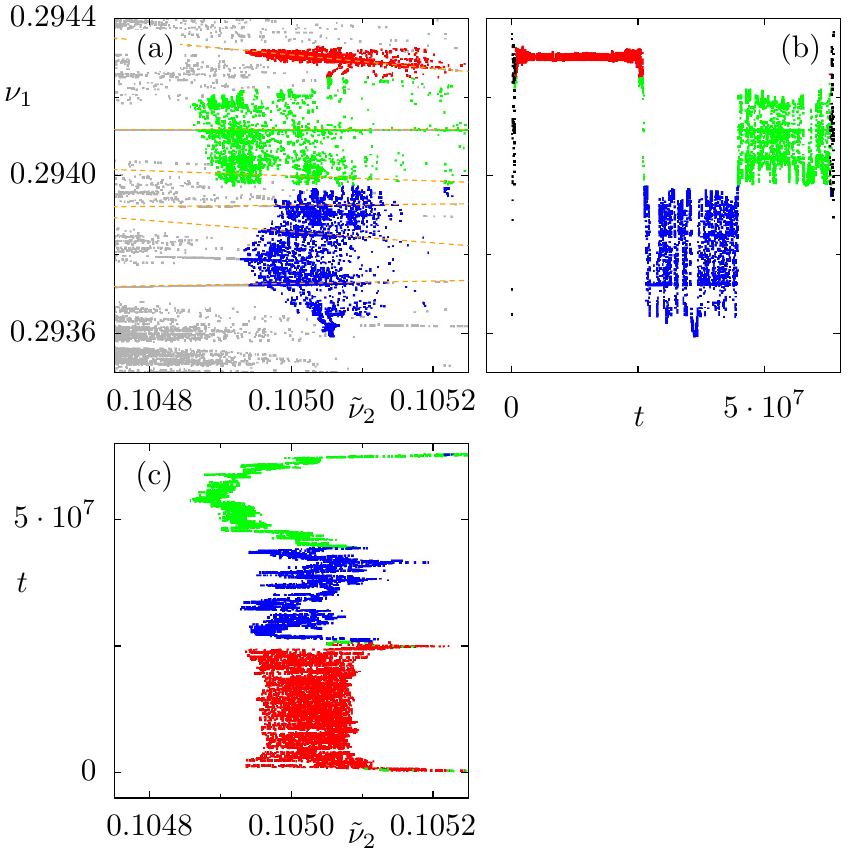}
\caption{\label{fig:time-frequency-4D} Transport across and along
  resonance channels for a single trapped orbit $t\in[0, 6.3\cdot
  10^7)$ displayed in local coordinates $(\nu_1, \tilde{\nu}_2)$ with
  $\tilde{\nu}_2=\nu_2+k\cdot (\nu_1-\nu_1^0)$, $k=1.94$,
  $\nu_1^0=0.294$. The boundary of (a) is indicated as a black box in
  Fig.~\ref{fig:freq-space-trapping}(b). Different frequency ranges in
  $\nu_1$ are colored red, green and blue. In (b), (c) the time
  dependence of $\nu_1$ (across) and $\tilde{\nu}_2$ (along) is shown
  (points outside of (a) in black).}
\end{figure}

In Fig.~\ref{fig:time-frequency-4D} this decomposition is demonstrated
for a trapped orbit by using convenient local coordinates $(\nu_1,
\tilde{\nu}_2)$. The boundary of Fig.~\ref{fig:time-frequency-4D}(a)
is shown as black box in Fig.~\ref{fig:freq-space-trapping}(a). The
transport across and along resonance channels is captured by
$\nu_1(t)$ in Fig.~\ref{fig:time-frequency-4D}(b) and
$\tilde{\nu}_2(t)$ in Fig.~\ref{fig:time-frequency-4D}(c),
respectively, as discussed below. Note that, while most trapped orbits
extend much more along $\nu_1$ than the example orbit, locally we
always observe the characteristics of
Fig.~\ref{fig:time-frequency-4D}.

{\it (iii.a) Signatures of partial barriers.}--- The transport across
resonance channels is captured in Fig.~\ref{fig:time-frequency-4D}(b).
The frequency $\nu_1(t)$ fluctuates within some frequency range over
long time intervals with sudden transitions to other ranges. In
\twoD{} maps such behavior is caused by partial
barriers~\cite{Lan2016}. Thus, Fig.~\ref{fig:time-frequency-4D}(b)
shows clear signatures
of partial barriers separating one or more resonance channels in a
generic \fourD{} map.

The origin of these partial barriers is a difficult open problem: Partial
barriers can be formed by the stable and unstable manifolds of the
family of hyperbolic \oneD{} tori, which is present in any resonance
channel~\cite{OnkLanKetBae2016} and studied in the context of normally
hyperbolic invariant manifolds~\cite{Wig1990}. Instead, we conjecture that the partial barriers are formed by
some families of cantori, as they
occur in Fig.~\ref{fig:time-frequency-4D}(a) in between resonance
channels. The existence of individual cantori in higher dimensions has
been shown in Ref.~\cite{MacMei1992}. The quantification of the flux across such a partial
barrier in higher-dimensional systems is conceptually problematic: As
the transport along the partial barrier is slow, only its local, and not its global, flux
is relevant for the dynamics of trapped orbits.
Note that in a special \fourD{} map, for which the relevant resonance channels are
parallel to the surface of the regular region, the transport across
resonance channels is the origin of the power-law
trapping~\cite{Lan2016}.

{\it (iii.b) Stochastic process with effective drift.}--- The
transport along a resonance channel is captured in
Fig.~\ref{fig:time-frequency-4D}(c). We conjecture that it can be
modeled by a one-dimensional stochastic process with an effective
drift: Such a drift was previously suggested due to the curvature of
the regular tori~\cite{KruKetKan2012, CasKal2015:p}. We propose that there is
a much stronger and more general effective drift $v$ due to the increase of the transversal \threeD{}
volume $A$ along the resonance channel: Assuming a fast transversal
diffusion one can project to a one-dimensional process along the
channel with a phase-space coordinate $x$. According to the Fick-Jacob
equation (Eq.~(2.6) in Ref.~\cite{KalPer2005}) this gives rise to a drift
\begin{equation}
\label{eq:fick-jacob}
v(x) = \frac{\partial_xA(x)}{A(x)} D(x)
\end{equation}
with the local
diffusion coefficient $D(x)$. The increase of the volume $A(x)$ along
a channel is already visible in the frequency space in
Fig.~\ref{fig:freq-space-trapping}(b) by the increasing widths of resonance channels
going towards the chaotic region.
Unfortunately, there is only weak numerical indication for such a
drift so far, as the slow transport along a channel is very difficult
to measure locally due to the lack of a sufficiently accurate
one-dimensional coordinate $x$~\cite{Lan2016}.

The transport along a resonance channel with an effective drift could
be one mechanism of the power-law trapping: Assuming a power-law
increase of the transversal volume along the channel $A(x) \sim
x^\alpha$ and using $D(x) \sim A(x)$~\cite{WooLicLie1990} the drift
is $v(x) = \delta x^{\alpha-1}$ with some factor $\delta$, according to
Eq.~\eqref{eq:fick-jacob}. For a process $x\in[0,x_{\text{abs}})$ with
an absorbing barrier at $x_{\text{abs}}$ one can derive a power-law decay
of the survival time distribution $P(t\gg 1) = t^{-\gamma}$ with
$\gamma = \frac{\delta-1}{\alpha-2}$, $\alpha>2$, $\delta>1$,
generalizing Ref.~\cite{MotMouGreKan2005} to $\delta \neq \alpha$.

{\it Outlook.}--- In this paper we advance towards understanding the
mechanism of power-law trapping in generic \fourD{} symplectic
maps.
The generality of the results is supported by similar observations we
have made for coupled twist maps~\cite{Lan2016} and a \threeD{}
billiard~\cite{Fir2014}. It remains to be shown how the two
local transport directions (across and along) of all the intersecting resonance channels
determine the global transport. For this, (iii.a) the origin of the
observed partial barriers should be clarified, i.e.\ whether they are
families of cantori, and (iii.b) the drift and diffusion along channels
should be quantified.
Apart from the power-law trapping, the effective
drift is crucial to understand the generic chaotic transport
in the Arnold web. Since all higher-dimensional systems share the crucial property
that regular tori are no barriers in phase space, the results obtained
for \fourD{} maps should be generalizable to even higher-dimensional
systems. \newline

We are grateful for discussions with J. D. Meiss, S. Keshavamurthy, J.
Laskar, E. G. Altmann, M. Richter, M. Toda, H. Kantz, and R. Klages.
Furthermore, we acknowledge support by the Deutsche
Forschungsgemeinschaft under grant KE~537/6--1. All \threeD{}
visualizations were created using \textsc{Mayavi}~\cite{RamVar2011}.

\end{document}